# Informed Consent:
# We Can Do Better to Defend Privacy

Dr. Frederik Zuiderveen Borgesius

Institute of Information Law, University of Amsterdam

f.j.zuiderveenborgesius [at] uva.nl

www.ivir.nl/medewerkerpagina/zuiderveenborgesius?lang=en

We need to rethink our approach to defend privacy on the internet. Currently, policymakers focus heavily on the idea of informed consent as a means to defend privacy. For instance, in many countries the law requires firms to obtain an individual's consent before they use data about her; with such informed consent requirements, the law aims to empower people to make privacy choices in their best interests. But behavioural studies cast doubt on this approach's effectiveness, as people tend to click OK to almost any request they see on their screens. To improve privacy protection, this article argues for a combined approach of protecting and empowering the individual.

This article discusses practical problems with informed consent as a means to protect privacy, and illustrates the problems with current data privacy rules regarding behavioural targeting. First, the privacy problems of behavioural targeting, and the central role of informed consent in privacy law are discussed. Following that, practical problems with informed consent are highlighted. Then, the article argues that policymakers should give more attention to rules that protect, rather than empower, people.

**Behavioural targeting and privacy**

Behavioural targeting is a marketing technique that involves tracking people's online behavior to use the collected information to display individually targeted Web advertisements to people.[1] Information captured for behavioural targeting can concern many online activities: what people read, which videos they watch, what they search for, etc. Individual profiles can be enriched with up-to-date location data of users of mobile devices, and other data that are gathered on and off line. Some providers of email or social network services analyse the contents of messages for marketing purposes. Vast amounts of information about hundreds of millions of people are collected for behavioural targeting.

In principle, behavioural targeting can benefit firms and consumers. Without explicitly paying with their money, people can enjoy access to online translation tools, online newspapers, and email accounts, and can watch videos and listen to music. Advertising supports many Internet services. However, advertising that doesn't require monitoring people's behaviour is also possible, such as contextual advertising: ads for cars on websites about cars. Therefore it's unclear whether behavioural targeting is needed to fund so-called "free" websites and services.

Behavioural targeting, however, raises privacy concerns. Three such concerns are: (i) chilling effects, (ii) a lack of control over personal information, and (iii) the risk of unfair discrimination and manipulation. First, behavioural targeting entails massive collection of information about people's online activities. Like other types of surveillance, this can cause chilling effects: people may adapt their behaviour if they suspect their activities may be monitored.[2]

---

[1] See on behavioural targeting: J Turow, *The Daily You: How the New Advertising Industry is Defining Your Identity and Your Worth* (New Haven, Yale University Press, 2011); Zuiderveen Borgesius FJ, Improving Privacy Protection in the Area of Behavioural Targeting (PhD thesis University of Amsterdam), Kluwer law International.

Second, people lack control over data concerning them. People do not know which information is collected, how it is used, and with whom it is shared. Large-scale personal data storage brings risks. For instance, a data breach could occur, or data could be used for unexpected purposes, such as identity fraud. Furthermore, the feeling of lost control is a privacy problem.

Third, behavioural targeting enables social sorting and discriminatory practices: firms can classify people as 'targets' and 'waste', and treat them accordingly. For instance, an advertiser could use discounts to lure affluent people to become regular customers – but might exclude poor people from the campaign.[3] And some fear that behavioural targeting could be used to manipulate people. Personalised advertising could become so effective that advertisers gain an unfair advantage over consumers.[4] Others worry that excessive personalization can lead to a filter bubble: If algorithms select the information we see on the web, we might each see a different image of the world.[5] Simply stated, the idea is that personalized advertising and other content could surreptitiously steer people's thoughts and actions. This fear seems most relevant when firms personalize not only ads but other content and services.

**Informed consent in data privacy law**

Informed consent and individual choice play a central role in many data privacy rules around the world. For instance, the OECD Privacy Guidelines say that personal data should be obtained "where appropriate, with the knowledge or consent of the data subject."[6] In the US, the consumer privacy bill of rights (in a White House report) emphasizes the importance of individual choice regarding how their data are used.[7] A body of Federal Trade Commission cases signal that the agency thinks that prominent notice and opt-in consent is necessary to collect sensitive personal information."[8]

---

[2] See Richards NM, 'Intellectual privacy' (2008) 87 Texas Law Review 387.
[3] J Turow, *The Daily You: How the New Advertising Industry is Defining Your Identity and Your Worth* (New Haven, Yale University Press, 2011).
[4] Calo MR. 'Digital market manipulation', George Washington Law Review, 82 George Washington Law Review 995 (2014).
[5] E Pariser, *The Filter Bubble* (London, Penguin Viking, 2011), p 9.

The EU e-Privacy Directive provides another example of the central role of informed consent in data privacy law. The directive requires any party that stores or accesses information on a user's device to obtain the user's informed consent.[9] The rule applies to many tracking technologies, including cookies. Valid consent requires a freely given, specific, and informed indication of wishes.[10] People can express their will in any form, but mere silence or inactivity isn't an expression of will. Some firms suggest they can presume "implied" consent if people don't block tracking cookies in their browser.[11] But this interpretation of the law seems incorrect. EU Data Protection Authorities say that the mere fact that a person leaves the browser settings untouched doesn't mean that she expressed her will to be tracked.[12]

Additionally, EU law requires consent to be voluntary ("freely given"): consent under too much pressure is not valid. Nevertheless, in most circumstances, current data privacy law will probably allow firms to offer take-it-or-leave-it choices. For example, website publishers are not prohibited from installing tracking walls that deny entry to visitors that don't consent to being tracked for behavioral targeting. But a tracking wall could make consent involuntary if people must use a website. For instance, according to the Dutch Data Protection Authority, the national public broadcasting organisation isn't allowed to use a tracking wall, because the only way to access certain information online is through the broadcaster's website.[13] EU Data Protection Authorities state that consent should be freely given, but do not say that current law prohibits tracking walls in all circumstances.[14]

---

[6] Collection Limitation Principle (nr. 7), OECD Guidelines Governing the Protection of Privacy and Transborder Flows of Personal Data.
[7] White House, 'Consumer Data Privacy in a Networked World: a Framework for Protecting Privacy and Promoting Innovation in the Global Digital Economy' (2012) <www.whitehouse.gov/sites/default/files/privacy-final.pdf>, p. 47; p. i.
[8] See: Federal Trade Commission, 'Protecting Consumer Privacy in an Era of Rapid Change: Recommendations for Businesses and Policymakers' (March 2012) www.ftc.gov/sites/default/files/documents/reports/federal-trade-commission-report-protecting-consumer-privacy-era-rapid-change-recommendations/120326privacyreport.pdf, p. 58 (hereinafter: FTC 2012).
[9] Directive 2002/58/EC as amended by Directive 2009/136/EC.
[10] Data Protection Directive 1995/46/EC.
[11] Interactive Advertising Bureau United Kingdom, 'Department for Business, Innovation & Skills consultation on implementing the revised EU electronic communications framework, IAB UK Response' (1 December 2012) www.iabuk.net/sites/default/files/IABUKresponsetoBISconsultationonimplementingtherevisedEUElectronicCommunicationsFramework_7427_0.pdf, p 2.
[12] Article 29 Working Party, 'Opinion 2/2010 on online behavioural advertising' (WP 171), 22 June 2010.

**Informed consent in practice**

Economic theory can help to analyse practical problems with consent to behavioural targeting. From an economic perspective, consenting to behavioural targeting can be seen as entering a market transaction with a firm. But this "transaction" is plagued by information asymmetries. Research shows that many people don't know to what extent their behaviour is tracked.[15] Therefore, their "choice" to disclose data in exchange for using a service can't be informed. Furthermore, data about people are collected, combined, and analyzed on a massive scale. A user might only leak scattered pieces of data, but firms could still construct detailed profiles by combining data from different sources

But, as behavioural economics research has highlighted, even if firms sought consent for behavioural targeting, information asymmetry would remain a problem.[16] People rarely know what a firm does with their personal data, and it's difficult to predict the consequences of future data usage. Firms have few incentives to offer privacy-friendly services because people can't assess whether a service is actually privacy-friendly. For instance, for users it is hard to detect whether their data are being captured during a website visit. Indeed, it seems websites don't use privacy as a competitive advantage: people are tracked for behavioral targeting on virtually every popular website.[17] There seems to be a comparable situation for smartphone apps.

---

[13] Dutch DPA, Letter to the State Secretary of Education, Culture and Science, 31 January 2013, www.cbpweb.nl/downloads_med/med_20130205-cookies-npo.pdf.
[14] Article 29 Working Party 2013, 'Working Document 02/2013 providing guidance on obtaining consent for cookies' (WP 208) 2 October 2013. See also FTC 2012, p. 52.
[15] See e.g. B Ur et al, 'Smart, Useful, Scary, Creepy: Perceptions of Online Behavioral Advertising' (Proceedings of the Eighth Symposium on Usable Privacy and Security ACM, 2012) 4.
[16] See A Acquisti and J Grossklags , 'What Can Behavioral Economics Teach Us About Privacy?' in A Acquisti et al (eds), *Digital Privacy: Theory, Technologies and Practices* (London, Auerbach Publications, Taylor and Francis Group 2007).

When animated by privacy-control rationales, data privacy law aims to reduce the information asymmetry. To illustrate, EU law requires firms to disclose certain information to individuals, for instance about the purposes for which personal data are used. Website publishers can use a privacy policy to comply with data protection law's transparency requirements.

However, the information asymmetry problem is hard to solve because of transaction costs for the individual, and again, information asymmetries regarding the meaning of privacy policies. Reading privacy policies would cost too much time, as they're often long, difficult to read, and vague. (One study calculated that it would take people several weeks per year if they read the privacy policy of every website they visit.[18]) Furthermore, privacy policies are too difficult for many. It's thus not surprising that almost nobody reads privacy policies. As a White House report puts it, "[o]nly in some fantasy world do users actually read these notices and understand their implications before clicking to indicate their consent."[19] In practice, data privacy law thus doesn't solve the information asymmetry problem.

To illustrate, a UK store obtained the soul of 7500 people. According to the website's terms and conditions, customers granted "a non transferable option to claim, for now and for ever more, your immortal soul," unless they opted out. By opting out, people could save their soul, and could receive a 5 pound

---

[17] T Vila, R Greenstadt and D Molnar, 'Why We Can't be Bothered to Read Privacy Policies. Models of Privacy Economics as a Lemons Market' in LJ Camp, and S. Lewis (eds), *Economics of Information Security* (Heidelberg, Springer, 2004).
[18] AM McDonald and LF Cranor, 'The Cost of Reading Privacy Policies' (2008) 4(3) *I/S: A Journal of Law and Policy for the Information Society* 540.
[19] White House (Podesta J et al.), 'Big Data: Seizing Opportunities, Preserving Values' (May 2014) www.whitehouse.gov/sites/default/files/docs/big_data_privacy_report_may_1_2014.pdf, p. xi.

voucher. But few people opted out. The firm later said it would not exercise its rights.[20]

Behavioural economics aims to improve the predictive power of economic theory, by including insights from psychology and behavioural studies. Behavioural economics suggests that people act structurally different than economic rational choice theory predicts. Because of their bounded rationality, people often rely on rules of thumb, or heuristics. Usually such mental shortcuts work fine, but they can also lead to behaviour that is not in people's self-interest. Several biases influence privacy choices, such as the status quo bias and the present bias.

An example of the status quo bias is that people are less likely to consent under an opt-in regime that requires an affirmative action for valid consent than under an opt-out regime in which people are assumed to consent if they don't object.[21] In this light, the continuous opt-in/opt-out discussion about behavioural targeting and other types of direct marketing concerns the question of who benefits from the status quo bias, the firm or the individual.

Present bias, or myopia, suggests that people often choose for immediate gratification and disregard future costs or disadvantages. For example, many find it hard to stick with a diet, or to save money. If a website has a tracking wall, and people can only use the site if they agree to behavioural targeting, they're likely to consent, thereby ignoring the costs of future privacy infringements.

In sum, behavioural economics shows that protecting privacy with the instrument of informed consent is fraught with problems. It's only a slight exaggeration to say: people don't read privacy policies; if they were to read, they wouldn't understand; if they understood, they wouldn't act. Additionally, if all competitors exploit information asymmetry and people's biases, a firm has to do the same to stay in business. Behavioural economics insights thus suggest that more regulatory intervention is justified in the area of behavioural targeting.

---

[20] Fox News, '7,500 Online Shoppers Unknowingly Sold Their Souls' (15 April 2010) www.foxnews.com/tech/2010/04/15/online-shoppers-unknowingly-sold-souls/.
[21] See Acquisti A and Gross R, 'Imagined Communities: Awareness, Information Sharing, and Privacy on the Facebook' (2006) 4258 6th International Workshop, PET 2006, Cambridge, UK, June 28-30, 2006 (Lecture Notes in Computer Science) 36.

**Individual protection, rather than empowerment**

Some data privacy laws contain rules that could defend privacy interests – also after somebody consents to processing. For instance, it follows from the OECD Privacy Guidelines that personal data should be protected by reasonable security safeguards.[22] Regulators in the EU and the US also emphasize the need for firms to secure the data they hold.[23]

And EU data privacy law has a stricter regime for "special categories of data", such as data revealing race, political opinions, health, or sex life.[24] In many EU countries, using special categories of personal data for direct marketing is prohibited; in other countries it is only allowed with the individuals' explicit consent.[25] Some firms target advertising based on categories such as "arthritis", "cardiovascular general health",[26] or "disabled/handicapped consumers."[27] Such firms process special categories of data. Strictly enforcing the existing rules on special categories of data could reduce privacy problems such as chilling effects. The rules on special categories of data could be interpreted in such a way that the collection context is taken into account. For example, tracking people's visits to websites with medical information should arguably be seen as processing "special categories of data," as the firm could infer health-related data from such tracking information.

Because the privacy risks involved in using health data for behavioural targeting seem to outweigh the possible societal benefits from allowing such practices, it should be considered to prohibit the use of any health related data for behavioural targeting, whether the individual gives explicit consent or not. The rules on special categories of data could be interpreted in such a way that the collection context is taken into account. For example, tracking people's visits to websites with medical information should arguably be seen as processing "special categories of data," as the firm could infer health-related data from such tracking information.

---

[22] Security Safeguards Principle (nr. 11), OECD Guidelines governing the protection of privacy and transborder flows of personal data.
[23] FTC 2012, p. 24-26. See also article 17 of the EU Data Protection Directive.
[24] Article 8 of the Data Protection Directive.
[25] There are exceptions, but these aren't relevant for behavioural targeting.
[26] Yahoo Privacy, 'All Standard Categories' http://info.yahoo.com/privacy/us/yahoo/opt_out/targeting/asc/details.html.
[27] Rocket Fuel, 'Health Related Segments' http://rocketfuel.com/downloads/Rocket%20Fuel%20Health%20Segments.pdf.

Because the privacy risks involved in using health data for behavioral targeting seem to outweigh the possible societal benefits from allowing such practices, policymakers should consider prohibiting the use of any health-related data for behavioral targeting, regardless of whether the individual gives explicit consent. A question that warrants more discussion is what the scope of such prohibitions should be. Should a prohibition of using health data for behavioral targeting also cover tracking daily visits to a website with gluten free recipes?

While strict enforcement of data privacy law's more protective principles could mitigate privacy problems somewhat, additional rules are probably needed. We saw that informed consent requirements, even if opt-in systems are required, won't be effective privacy nudges as long as firms are allowed to offer take-it-or-leave-it choices.

**Broadening the debate**

It is time to extend the privacy debate beyond informed consent. Aiming for transparency and consent will not suffice to ensure a reasonable level of privacy. Consumer law illustrates how empowerment and protection rules can both be used, as complementary tools. In many circumstances, consumer law requires firms to disclose information to consumers (calories, delivery costs…). Such transparency requirements aim to empower consumers to make decisions according to their preferences. Other rules in consumer law aim to protect consumers. For instance, some food ingredients may not be used at all, and for many products there are minimum safety standards.

Some have suggested that policymakers should focus more on data use, and less on informed consent for data collection.[28] Focusing mostly on data use, however, has considerable risks. I argue strongly against leaving collection mostly unregulated. Many privacy problems, such as chilling effects, occur already because of data collection. Apart from that, in Europe a regime that leaves collection unregulated would be difficult to reconcile with fundamental rights case law and treaties.[29]

What should policymakers do about take-it-or-leave-it choices such as tracking walls? The law could prohibit tracking walls in certain circumstances. For instance, public service broadcasters often receive public funding and play a special role in educating and informing the public, and in promoting the values of democratic societies. But if people fear their behaviour is being monitored, they might forego using public service media. Therefore, policymakers should prohibit public service broadcasters from installing tracking walls on their websites. Policymakers could also go one step further, and prohibit all third party tracking for behavioural targeting on public service media.

More generally it's questionable whether it's appropriate for websites of state bodies to allow third party tracking for behavioural targeting – even when people consent. In practice, public sector websites might use third party widgets such as social media buttons; website publishers might not realise that such widgets may expose visitors to privacy-invasive tracking. But it's not evident why the public sector should facilitate tracking people's behaviour for commercial purposes. Therefore, policymakers should consider prohibiting all tracking for behavioural targeting on public sector websites.

Policymakers can add prohibitions to their toolbox to regulate behavioural targeting, though it would be difficult to define prohibitions so that they're not over- or underinclusive. Difficult questions are ahead for researchers and policymakers. A careful balance must be struck between undue paternalism and protecting people. The legal protection of privacy will remain a learning process. If new rules were adopted, their practical effect would have to be evaluated. The problems with the current informed consent requirements demonstrate that regulation that looks good on paper may not effectively protect privacy in practice.

---

[28] See for instance President's Council of Advisors on Science and Technology, "Big Data and Privacy: A Technological Perspective," 2014, and Craig Mundie, "Privacy Pragmatism: Focus on Data Use, Not Data Collection," Foreign Affairs.

[29] See for instance: European Court of Human Rights, S. and Marper v. United Kingdom, No. 30562/04 and 30566/04. 4 December 2008, par. 67; Court of Justice of the European Union, C-293/12 and C-594/12, Digital Rights Ireland Ltd, 8 April 2014, par. 29.

**Defending privacy with technology**

The distinction between empowerment and protection rules in the law could also inform discussions about technical privacy defence tools. Empowering users is an important goal. For example, technology might help to foster meaningful transparency regarding data processing and profiling. And user-friendly mechanisms are needed to give, withhold, or retract consent. However, in some circumstances people may benefit more from protection against risks, than from being confronted with transparency and choices. Examples of more protective technical approaches might include services that automatically secure personal information, metadata or communications, regardless of the user's initiative.

**Conclusion**

In conclusion, there's no silver bullet to improve privacy protection in the area of behavioural targeting. While current regulation often emphasises individual empowerment, without much reflection on practical issues, a combined approach could be used to protect and empower people. To improve privacy protection, current data privacy law should be more strictly enforced. But the limited potential of informed consent as a privacy protection measure should be taken into account.

\* \* \*

**Acknowledgements**

This article is based on the author's PhD thesis: Zuiderveen Borgesius FJ, Improving Privacy Protection in the Area of Behavioural Targeting (PhD thesis University of Amsterdam), Kluwer law International.

**About the author**

Frederik Zuiderveen Borgesius is a researcher at the Institute for Information Law (University of Amsterdam) www.ivir.nl/medewerkerpagina/zuiderveenborgesius?lang=en